\documentclass[useAMS,preprint2]{aastex}

\newcommand{\teff}{\mbox{$T_{\rm eff}$}}
\newcommand{\logg}{\mbox{$\log g$}}
\newcommand{\vsini}{\mbox{$v \sin i$}}

\newcommand{\kms}{\mbox{km\,s$^{-1}$}}

\newcommand{\Teff}{\mbox{T$_{\rm eff}$}}
\newcommand{\rhostar}{\mbox{${\rho}_{\star}$}}
\newcommand{\taustar}{\mbox{${\tau}_{\star}$}}
\newcommand{\etal}{et\,al.}
\newcommand{\kmps}{km\,s$^{-1}$}

\newcommand{\Msolar}{\mbox{M$_{\sun}$}}

\newcommand{\Mstar}{\mbox{${M}_{\star}$}}
\newcommand{\sqiglt}{\hbox{\rlap{\lower.55ex \hbox {$\sim$}}
	\kern-.3em \raise.4ex \hbox{$<$}\,}}
\newcommand{\sqiggt}{\hbox{\rlap{\lower.55ex \hbox {$\sim$}}
	\kern-.3em \raise.4ex \hbox{$>$}\,}}

\begin{document}

\shorttitle{Three transiting exoplanets}

\shortauthors{Hellier et al}
\title{Three WASP-South transiting exoplanets: \\ WASP-74b, WASP-83b \&\ WASP-89b}

\author{Coel Hellier\altaffilmark{1}, 
D.R. Anderson\altaffilmark{1}, 
A. Collier Cameron\altaffilmark{2}, 
L. Delrez\altaffilmark{3},
M. Gillon\altaffilmark{3},
E. Jehin\altaffilmark{3}, 
M. Lendl\altaffilmark{3,4}, 
P.F.L. Maxted\altaffilmark{1}, 
F. Pepe\altaffilmark{4}, 
D. Pollacco\altaffilmark{5}, 
D. Queloz\altaffilmark{4,6},
D. S\'egransan\altaffilmark{4}, 
B. Smalley\altaffilmark{1}, 
A.M.S. Smith\altaffilmark{1,7}, 
J. Southworth\altaffilmark{1},
A.H.M.J. Triaud\altaffilmark{4,8},
O.D. Turner\altaffilmark{1}, 
S. Udry\altaffilmark{4}, \\ \&\ R.G. West\altaffilmark{5}}

\altaffiltext{1}{Astrophysics Group, Keele University, Staffordshire, ST5 5BG, UK}
\altaffiltext{2}{SUPA, School of Physics and Astronomy, University of St.\ Andrews, North Haugh,  Fife, KY16 9SS, UK}
\altaffiltext{3}{Institut d'Astrophysique et de G\'eophysique, Universit\'e, Li\`ege, All\'ee du 6 Ao\^ut, 17, Bat. B5C, Li\`ege 1, Belgium}
\altaffiltext{4}{Observatoire astronomique de l'Universit\'e de Gen\`e, 51 ch. des Maillettes, 1290 Sauverny, Switzerland}
\altaffiltext{5}{Department of Physics, University of Warwick, Gibbet Hill Road, Coventry CV4 7AL, UK}
\altaffiltext{6}{Cavendish Laboratory, J J Thomson Avenue, Cambridge, CB3 0HE, UK}
\altaffiltext{7}{N.~Copernicus Astronomical Centre, Polish Academy of Sciences, Bartycka 18, 00-716 Warsaw, Poland}
\altaffiltext{8}{Department of Physics and Kavli Institute for Astrophysics \&\ Space Research, Massachusetts Institute of Technology, Cambridge, MA 02139, USA}


\begin{abstract}
We report the discovery of three new transiting hot Jupiters by WASP-South together with the TRAPPIST photometer and the Euler/CORALIE spectrograph.  

WASP-74b orbits a star of $V$ = 9.7, making it one of the brighter systems accessible to Southern telescopes. It is a 0.95  M$_{\rm Jup}$ planet with a moderately bloated radius of 1.5 R$_{\rm Jup}$ in a 2-d orbit around a slightly evolved F9 star.  

WASP-83b is a Saturn-mass planet at  0.3 M$_{\rm Jup}$ with a radius of 1.0  R$_{\rm Jup}$. It is in a 5-d orbit around a fainter ($V$ = 12.9) G8 star.

WASP-89b is a 6 M$_{\rm Jup}$ planet in a 3-d orbit with an eccentricity of $e$ = 0.2.  It is thus similar to massive, eccentric planets such as XO-3b and HAT-P-2b, except that those planets orbit F stars whereas WASP-89 is a K star. The $V$ = 13.1 host star is magnetically active, showing a rotation period of 20.2 d, while star spots are visible in the transits.  There are indications that the planet's orbit is aligned with the stellar spin.  WASP-89 is a good target for an extensive study of transits of star spots. 
\end{abstract}


\section{Introduction}
The combination of the WASP-South survey instrument, the Euler/CORALIE spectrograph and the robotic  TRAPPIST photometer continue to be an efficient team for the discovery of  transiting exoplanets around stars of $V$ = 9--13 in the Southern Hemisphere (e.g.~Hellier \etal\ 2014; Anderson \etal\ 2014a).  Ongoing discoveries are important for expanding our census of the hot-Jupiter population, while exoplanets transiting relatively bright stars are good targets for followup studies. In this paper we report three new discoveries: WASP-74b, which orbits a bright $V$ = 9.7 star; WASP-83b, a more moderately bloated Saturn-mass planet, which, with a period of 4.97 d demonstrates the capability of a single-longitude transit search to find planets with integer-day periods; and WASP-89b, a massive planet in a short and eccentric orbit around a magnetically active K star.   

\section{Observations}
The observational and analysis techniques used here are similar to those in recent WASP-South discovery papers  (e.g.\ Hellier \etal\ 2012; Anderson \etal\ 2014b), and so are reported briefly. WASP-South surveys the Southern sky using an array of 200mm f/1.8 lenses and a cadence of $\sim$\,10 mins (see Pollacco \etal\ 2006).  Transit searching of accumulated lightcurves (Collier-Cameron \etal\ 2007a) leads to tens of thousands of possible candidates, of which the vast majority are false alarms resulting from the limitations of the photometry.  The best 1 per cent are selected by eye as candidates and are passed to TRAPPIST (a robotic 0.6-m photometric telescope) and to the 1.2-m Euler/CORALIE spectrograph (for radial-velocity observations).     About 1 in 12 candidates 
turns out to be a planet, with most of the others being astrophysical transit mimics (blended or grazing eclipsing binaries).  Higher-quality transit lightcurves are then obtained with TRAPPIST (Jehin \etal\ 2011) and with EulerCAM (Lendl \etal\ 2012). We have also observed a transit of WASP-74b using RISE on the Liverpool Telescope (see Steele \etal\ 2008). 

  A list of the observations reported here is given in Table~1 while the CORALIE radial velocities are listed in Table~A1.  

\begin{table}[t]
\caption{Observations\protect\rule[-1.5mm]{0mm}{2mm}}  
\begin{tabular}{lcr}
\hline 
Facility & Date &  \\ [0.5mm] \hline
\multicolumn{3}{l}{{\bf WASP-74:}}\\  
WASP-South & 2010 May--2012 Jun & 10\,000 points \\
CORALIE  & 2011 Aug--2012 Oct  &   20 RVs \\
EulerCAM  & 2012 May 07 & Gunn $r$ filter \\ 
TRAPPIST & 2012 May 07 & $z^{\prime}$ band \\
EulerCAM  & 2012 May 22 & Gunn $r$ filter \\ 
TRAPPIST & 2012 May 22 & $z^{\prime}$ band \\
TRAPPIST & 2012 Jun 21 & $z^{\prime}$ band \\
TRAPPIST & 2012 Jun 23 & $z^{\prime}$ band \\
TRAPPIST & 2012 Sep 04 & $z^{\prime}$ band \\
TRAPPIST & 2013 Jun 27 & $I+z$ band \\ 
LT/RISE & 2014 Aug 19 & $V+R$ \\ [0.5mm] 
\multicolumn{3}{l}{\bf WASP-83:}\\  
WASP-South & 2006 May--2010 Jun & 20\,600 points \\
CORALIE  & 2011 Mar--2013 Feb  &   28 RVs \\
TRAPPIST & 2012 Jan 07 & Blue-block filter \\
TRAPPIST & 2012 Jan 22 & Blue-block filter \\
TRAPPIST & 2012 Feb 06 & Blue-block filter \\ [0.5mm]
\multicolumn{3}{l}{\bf WASP-89:}\\  
WASP-South & 2008 Jun--2012 Jun  & 18\,000 points \\
CORALIE  & 2011 May--2013 May  &   20 RVs \\
TRAPPIST & 2012 Aug 26 & Blue-block filter \\
EulerCAM & 2012 Sep 12 & Gunn $r$ filter \\ 
TRAPPIST & 2012 Sep 12 & Blue-block filter \\
EulerCAM & 2012 Oct 02 & Gunn $r$ filter \\ 
TRAPPIST & 2012 Oct 02 & Blue-block filter \\
TRAPPIST & 2013 Jun 14 & Blue-block filter \\
TRAPPIST & 2013 Aug 27 & Blue-block filter \\ [0.5mm] \hline
\end{tabular} 
\end{table}

\section{The host stars} 
We used the CORALIE spectra to analyse the three host stars, co-adding
the standard pipeline reduction products to produce spectra with S/N 
ratios of  150:1, 100:1 and 30:1 for  WASP-74, WASP-83 and WASP-89 respectively.  Our analysis methods are described in Doyle \etal\ (2013).  The  effective temperature (\teff) estimate comes from the excitation balance of Fe~{\sc i} lines, while the surface gravity (\logg) estimates comes from the
ionisation balance of Fe~{\sc i} and Fe~{\sc ii} and the Ca~{\sc i} line
at 6439{\AA} and the Na~{\sc i} D lines. The metallicity was determined from
equivalent width measurements of several unblended lines. The quoted
error estimates include that given by the uncertainties in \teff\ and
\logg, as well as the scatter due to measurement and atomic data
uncertainties.

The projected stellar rotation velocity (\vsini) was determined by
fitting the profiles of several unblended Fe~{\sc i} lines. Values of
macroturbulent velocity of 3.9 $\pm$ 0.7 {\kms} and 2.9 $\pm$ 0.7 {\kms}
were adopted for WASP-74 and WASP-83, using the calibration of Doyle
\etal\ (2014). For WASP-89, however, macroturbulence was assumed to be
zero, since for mid-K stars it is expected to be lower than that of
thermal broadening (Gray 2008). 

The parameters obtained from the analysis are given in Tables 2 to
4. The quoted spectral type derives from \teff, using the values in
Gray (2008). Abundances are relative to the solar values obtained by
Asplund \etal\ (2009). Gyrochronological age estimates derive from the
measured \vsini, assuming that the star's spin is perpendicular to us,
so that this is the true equatorial speed.  This is then combined with
the stellar radius to give a rotational period, to compare with the
values in Barnes (2007). Lithium age estimates come from values in Sestito \&\ Randich (2005). We also list proper motions from the
UCAC4 catalogue of Zacharias \etal\ (2013).

We searched the WASP photometry of each star for rotational
modulations by using a sine-wave fitting algorithm as described by
Maxted \etal\ (2011). We estimated the significance of periodicities
by subtracting the fitted transit lightcurve and then repeatedly and
randomly permuting the nights of observation.  We found a significant 
modulation in WASP-89 (see Section 8.1) but not in the other two stars.

\section{System parameters}
The CORALIE radial-velocity measurements were combined with the WASP,
EulerCAM and TRAPPIST photometry in a simultaneous Markov-chain
Monte-Carlo (MCMC) analysis to find the system parameters.
For details of our methods see Collier Cameron \etal\ (2007b). 
The limb-darkening parameters are noted in each Table, and are
taken from the 4-parameter non-linear law of Claret (2000).

For WASP-89b the orbital eccentricity is significant and was fitted as
a free parameter.  For WASP-74b and WASP-83b we imposed a circular
orbit during the analysis (see Anderson \etal\ 2012 for the rationale
for this).

\begin{table}
\caption{System parameters for WASP-74.}  
\begin{tabular}{lc}
\multicolumn{2}{l}{1SWASP\,J201809.32--010432.6}\\
\multicolumn{2}{l}{2MASS\,20180931--0104324}\\
\multicolumn{2}{l}{RA\,=\,20$^{\rm h}$18$^{\rm m}09.32$$^{\rm s}$, 
Dec\,=\,--01$^{\circ}$04$^{'}$32.6$^{''}$ (J2000)}\\
\multicolumn{2}{l}{$V$ mag = 9.7}  \\ 
\multicolumn{2}{l}{Rotational modulation\ \ \ $<$\,0.7 mmag (95\%)}\\
\multicolumn{2}{l}{pm (RA) 1.6\,$\pm$\,1.0 (Dec) --64.3\,$\pm$\,0.7 mas/yr}\\
\hline
\multicolumn{2}{l}{Stellar parameters from spectroscopic analysis.\rule[-1.5mm]{
0mm}{2mm}} \\ \hline 
Spectral type & F9 \\
$T_{\rm eff}$ (K)  & 5990  $\pm$ 110  \\
$\log g$      & 4.39 $\pm$ 0.07    \\
$v\,\sin I$ (km\,s$^{-1}$)     &    4.1 $\pm$ 0.8      \\
{[Fe/H]}   &   +0.39 $\pm$ 0.13     \\
log A(Li)  &    2.74 $\pm$ 0.09      \\
Age (Lithium) [Gy]  &    0.5\,$\sim$\,2          \\
Age (Gyro) [Gy]     &   $2.0^{+1.6}_{-1.0}$  \\ 
Distance [pc]  & 120 $\pm$ 20  \\ \hline 
\multicolumn{2}{l}{Parameters from MCMC analysis.\rule[-1.5mm]{0mm}{3mm}} \\
\hline 
$P$ (d) &    2.137750   $\pm$ 0.000001 \\
$T_{\rm c}$ (HJD)\,(UTC) & 245\,6506.8918 $\pm$ 0.0002\\ 
$T_{\rm 14}$ (d) & 0.0955 $\pm$ 0.0008\\ 
$T_{\rm 12}=T_{\rm 34}$ (d) & 0.0288  $\pm$ 0.0014  \\
$\Delta F=R_{\rm P}^{2}$/R$_{*}^{2}$ &  0.00961  $\pm$ 0.00014\\ 
$b$ & 0.860 $\pm$ 0.006 \\
$i$ ($^\circ$) & 79.81 $\pm$ 0.24 \\
$K_{\rm 1}$ (km s$^{-1}$) &  0.1141 $\pm$ 0.0014\\ 
$\gamma$ (km s$^{-1}$) & --15.767  $\pm$ 0.001\\ 
$e$ & 0 (adopted) ($<$\,0.07 at 3$\sigma$) \\ 
$M_{\rm *}$ (M$_{\rm \odot}$) &  1.48 $\pm$ 0.12\\ 
$R_{\rm *}$ (R$_{\rm \odot}$) &  1.64  $\pm$ 0.05\\
$\log g_{*}$ (cgs) & 4.180 $\pm$ 0.018 \\
$\rho_{\rm *}$ ($\rho_{\rm \odot}$) &   0.338 $\pm$ 0.018\\
$T_{\rm eff}$ (K) & 5970 $\pm$ 110\\
$M_{\rm P}$ (M$_{\rm Jup}$) & 0.95 $\pm$ 0.06\\
$R_{\rm P}$ (R$_{\rm Jup}$) &  1.56 $\pm$ 0.06\\
$\log g_{\rm P}$ (cgs) & 2.95 $\pm$ 0.02 \\
$\rho_{\rm P}$ ($\rho_{\rm J}$) & 0.25 $\pm$ 0.02\\
$a$ (AU)  & 0.037 $\pm$ 0.001\\
$T_{\rm P, A=0}$ (K) & 1910 $\pm$ 40\\ [0.5mm] \hline 
\multicolumn{2}{l}{Errors are 1$\sigma$; Limb-darkening coefficients were:}\\
\multicolumn{2}{l}{{\small Trap\,$z$: a1 = 0.757, a2 = --0.591, a3 = 0.890, 
a4 = --0.416}}\\ 
\multicolumn{2}{l}{{\small Euler \&\ RISE: a1 =   0.669 , a2 = --0.284 , a3 = 0.765, a4 = --0.395}}\\ \hline
\end{tabular} 
\end{table}

\begin{figure}
\hspace*{-12mm}\includegraphics[width=10cm]{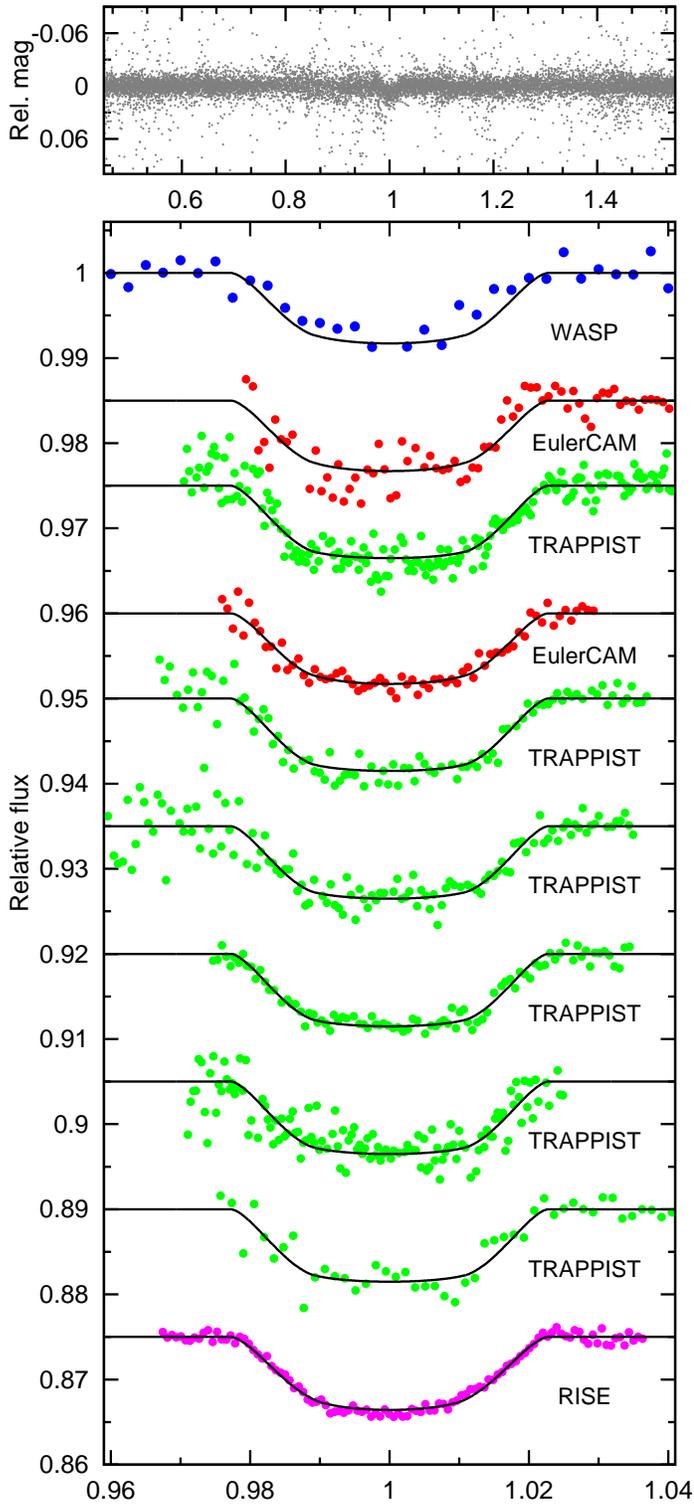}\\ [-2mm]
\caption{WASP-74b discovery photometry: (Top) The WASP data folded on the 
transit period. (Second panel) The binned WASP data with (offset) the
follow-up transit lightcurves (ordered from the top as 
in Table~1) together with the 
fitted MCMC model.   The two EulerCAM lightcurves are of the same transit as the TRAPPIST data directly below them.}
\end{figure}

\begin{figure}
\hspace*{-5mm}\includegraphics[width=10cm]{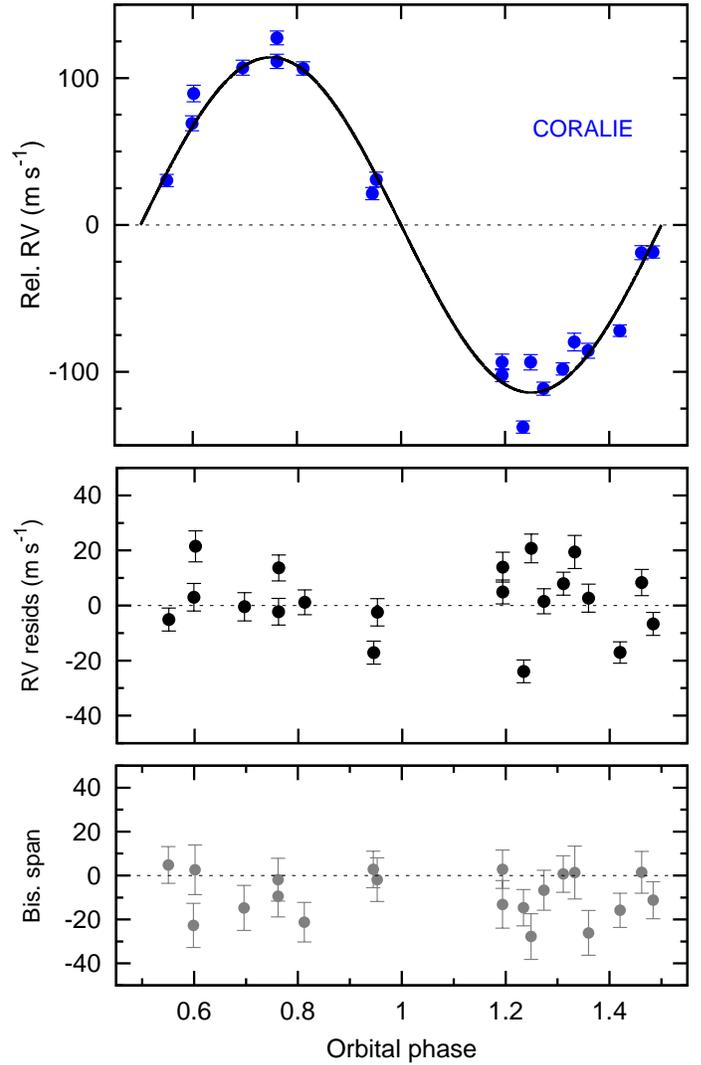}\\ [-2mm]
\caption{WASP-74b radial velocities and fitted model (top) along with (middle) the residuals and (bottom) the bisector spans; the absence of any correlation with radial velocity is a check against transit mimics.}
\end{figure}

\begin{table}
\caption{System parameters for WASP-83.}  
\begin{tabular}{lc}
\multicolumn{2}{l}{1SWASP\,J124036.51--191703.4}\\
\multicolumn{2}{l}{2MASS\,12403650--1917032}\\
\multicolumn{2}{l}{RA\,=\,12$^{\rm h}$40$^{\rm m}36.51$$^{\rm s}$, 
Dec\,=\,--19$^{\circ}$17$^{'}$03.4$^{''}$ (J2000)}\\
\multicolumn{2}{l}{$V$ mag = 12.9}  \\ 
\multicolumn{2}{l}{Rotational modulation\ \ \ $<$\,1.5 mmag (95\%)}\\
\multicolumn{2}{l}{pm (RA) --21.6\,$\pm$\,1.9 (Dec) --11.5\,$\pm$\,2.1 mas/yr}\\
\hline
\multicolumn{2}{l}{Stellar parameters from spectroscopic analysis.\rule[-1.5mm]{
0mm}{2mm}} \\ \hline 
Spectral type & G8 \\
$T_{\rm eff}$ (K)  & 5480  $\pm$ 110  \\
$\log g$      & 4.34 $\pm$ 0.08    \\
$v\,\sin I$ (km\,s$^{-1}$)     &    $<$ 0.5   \\
{[Fe/H]}   &   +0.29 $\pm$ 0.12     \\
log A(Li)  &     $<$0.75     \\
Age (Lithium) [Gy]  &    $\ga$ 5          \\
Distance [pc]  & 300 $\pm$ 50   \\ \hline
\multicolumn{2}{l}{Parameters from MCMC analysis.\rule[-1.5mm]{0mm}{3mm}} \\
\hline 
$P$ (d) &    4.971252   $\pm$ 0.000015 \\
$T_{\rm c}$ (HJD)\,(UTC) & 245\,5928.8853 $\pm$ 0.0004\\ 
$T_{\rm 14}$ (d) & 0.1402 $\pm$ 0.0015\\ 
$T_{\rm 12}=T_{\rm 34}$ (d) & 0.0136  $\pm$ 0.0017  \\
$\Delta F=R_{\rm P}^{2}$/R$_{*}^{2}$ & 0.0104   $\pm$ 0.0004\\ 
$b$ & 0.23 $\pm$ 0.15 \\
$i$ ($^\circ$) & 88.9 $\pm$ 0.7 \\
$K_{\rm 1}$ (km s$^{-1}$) &  0.0329 $\pm$ 0.0031\\ 
$\gamma$ (km s$^{-1}$) & 31.549  $\pm$ 0.002\\ 
$e$ & 0 (adopted) ($<$\,0.3 at 3$\sigma$) \\ 
$M_{\rm *}$ (M$_{\rm \odot}$) &  1.11 $\pm$ 0.09\\ 
$R_{\rm *}$ (R$_{\rm \odot}$) &  1.05  $^{+0.06}_{-0.04}$\\
$\log g_{*}$ (cgs) & 4.44 $^{+0.02}_{-0.04}$  \\
$\rho_{\rm *}$ ($\rho_{\rm \odot}$) &   0.97 $^{+0.07}_{-0.13}$ \\
$T_{\rm eff}$ (K) & 5510 $\pm$ 110\\
$M_{\rm P}$ (M$_{\rm Jup}$) & 0.30 $\pm$ 0.03\\
$R_{\rm P}$ (R$_{\rm Jup}$) &  1.04 $^{+0.08}_{-0.05}$\\
$\log g_{\rm P}$ (cgs) & 2.79 $\pm$ 0.06 \\
$\rho_{\rm P}$ ($\rho_{\rm J}$) & 0.26 $\pm$ 0.05\\
$a$ (AU)  & 0.059 $\pm$ 0.001\\
$T_{\rm P, A=0}$ (K) & 1120 $\pm$ 30\\ [0.5mm] \hline 
\multicolumn{2}{l}{Errors are 1$\sigma$; Limb-darkening coefficients were:}\\
\multicolumn{2}{l}{{\small a1 =  0.747, a2 = --0.649, a3 = 1.277, 
a4 = --0.584}}\\ \hline
\end{tabular} 
\end{table}

\begin{figure}
\hspace*{-10mm}\includegraphics[width=10cm]{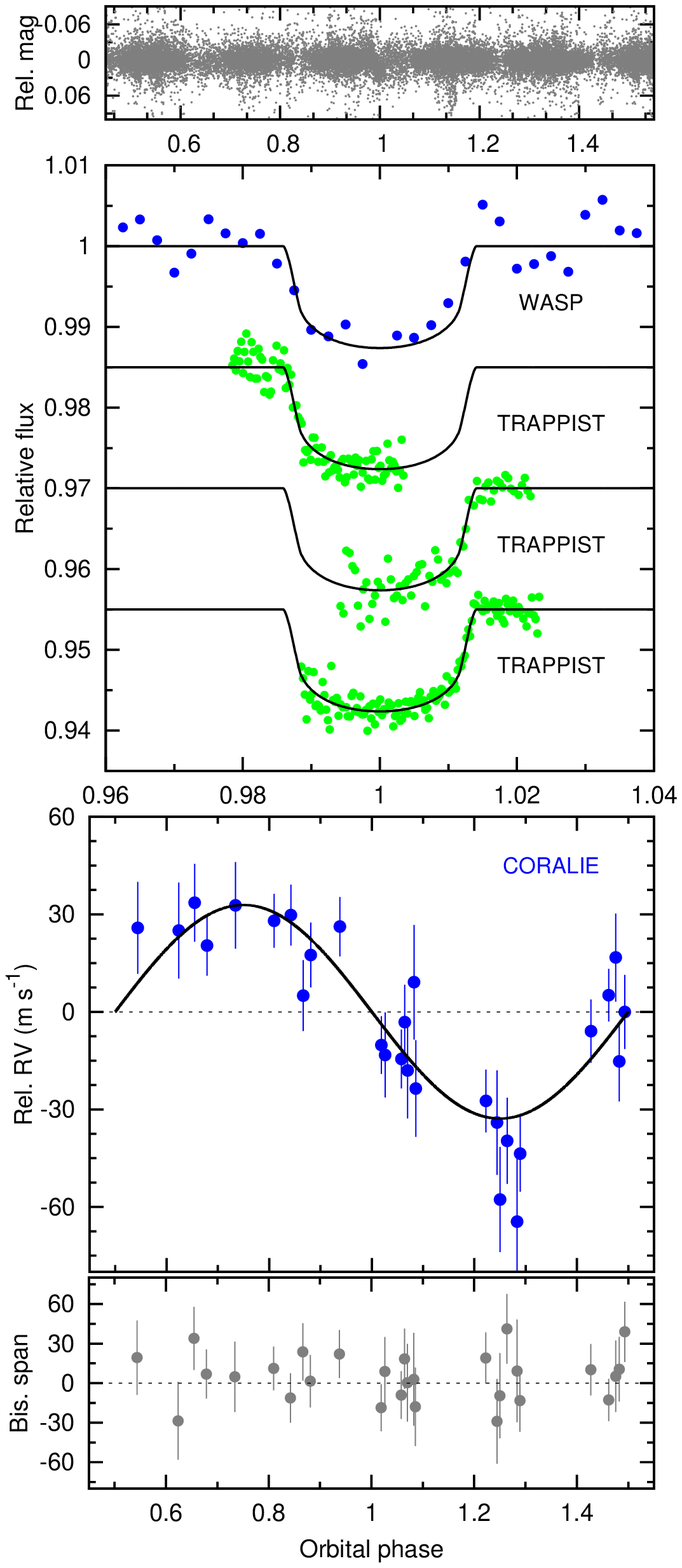}\\ [-2mm]
\caption{WASP-83b discovery data, as for Figs.~1 and 2.}
\end{figure}

\begin{table}
\caption{System parameters for WASP-89.}  
\begin{tabular}{lc}
\multicolumn{2}{l}{1SWASP\,J205535.98--185816.1}\\
\multicolumn{2}{l}{2MASS\,20553599--1858159}\\
\multicolumn{2}{l}{RA\,=\,20$^{\rm h}$55$^{\rm m}35.98$$^{\rm s}$, 
Dec\,=\,--18$^{\circ}$58$^{'}$16.1$^{''}$ (J2000)}\\
\multicolumn{2}{l}{$V$ mag = 13.1}  \\ 
\multicolumn{2}{l}{pm (RA) 13.7\,$\pm$\,1.5 (Dec) --63.0\,$\pm$\,1.4 mas/yr}\\
\hline
\multicolumn{2}{l}{Stellar parameters from spectroscopic analysis.\rule[-1.5mm]{
0mm}{2mm}} \\ \hline 
Spectral type & K3 \\
$T_{\rm eff}$ (K)  & 4955  $\pm$ 100  \\
$\log g$      & 4.31 $\pm$ 0.16    \\
$v\,\sin I$ (km\,s$^{-1}$)     &    2.5 $\pm$ 0.9      \\
{[Fe/H]}   &   +0.15 $\pm$ 0.14     \\
log A(Li)  &   $<$ 0.24       \\
Age (Lithium) [Gy]  &   $\ga$ 0.5          \\
Age (Gyro) [Gy]     &  $1.3^{+1.5}_{-0.8}$     \\ \hline 
\multicolumn{2}{l}{Parameters from MCMC analysis.\rule[-1.5mm]{0mm}{3mm}} \\
\hline 
$P$ (d) &   3.3564227    $\pm$ 0.0000025 \\
$T_{\rm c}$ (HJD)\,(UTC) & 245\,6207.02114 $\pm$ 0.00012\\ 
$T_{\rm 14}$ (d) & 0.1025 $\pm$ 0.0004\\ 
$T_{\rm 12}=T_{\rm 34}$ (d) & 0.0112  $\pm$ 0.0003  \\
$\Delta F=R_{\rm P}^{2}$/R$_{*}^{2}$ &  0.0149  $\pm$ 0.0002\\ 
$b$ & 0.10 $\pm$ 0.08 \\
$i$ ($^\circ$) & 89.4 $\pm$ 0.5 \\
$K_{\rm 1}$ (km s$^{-1}$) &  0.848 $\pm$ 0.013\\ 
$\gamma$ (km s$^{-1}$) &  21.088 $\pm$ 0.008\\ 
$e$ & 0.193  $\pm$ 0.009 \\
$\omega$ (deg) & 28 $\pm$ 4 \\
$M_{\rm *}$ (M$_{\rm \odot}$) &  0.92 $\pm$ 0.08\\ 
$R_{\rm *}$ (R$_{\rm \odot}$) &  0.88 $\pm$ 0.03\\
$\log g_{*}$ (cgs) & 4.515 $\pm$ 0.018\\ 
$\rho_{\rm *}$ ($\rho_{\rm \odot}$) &   1.36 $\pm$0.07 \\
$T_{\rm eff}$ (K) & 5130 $\pm$ 90\\
$M_{\rm P}$ (M$_{\rm Jup}$) & 5.9 $\pm$ 0.4\\
$R_{\rm P}$ (R$_{\rm Jup}$) &  1.04 $\pm$ 0.04\\ 
$\log g_{\rm P}$ (cgs) & 4.10 $\pm$ 0.02 \\
$\rho_{\rm P}$ ($\rho_{\rm J}$) & 5.27 $\pm$ 0.33\\
$a$ (AU)  & 0.0427 $\pm$ 0.0012\\
$T_{\rm P, A=0}$ (K) & 1120 $\pm$ 20\\ [0.5mm] \hline 
\multicolumn{2}{l}{Errors are 1$\sigma$; Limb-darkening coefficients were:}\\
\multicolumn{2}{l}{{\small a1 =  0.741, a2 = --0.739, a3 = 1.427, 
a4 = --0.620}}\\ \hline
\end{tabular} 
\end{table}

\begin{figure}
\hspace*{-10mm}\includegraphics[width=10cm]{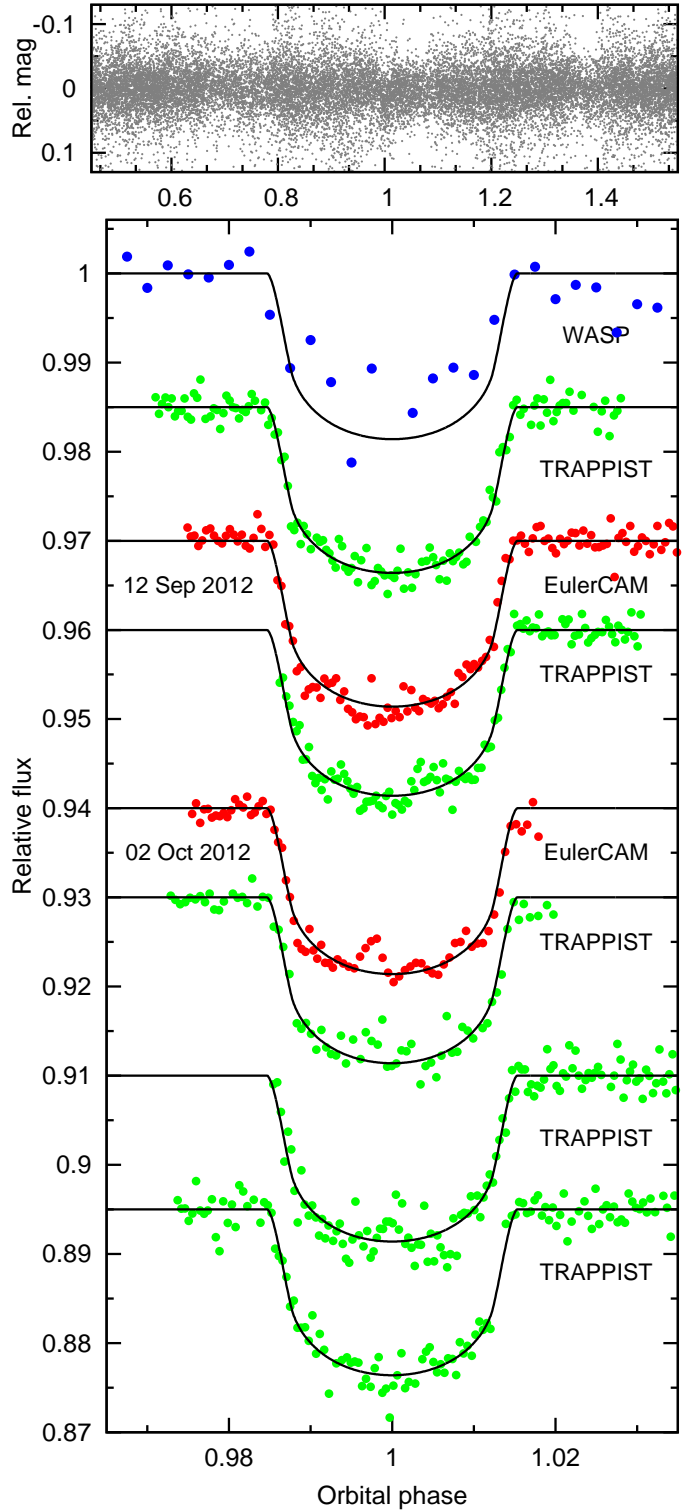}\\ [-2mm]
\caption{WASP-89b discovery data.  The Euler and TRAPPIST observations on 12 Sep 2012 and on 02 Oct 2012 are of the same transit, and hence these transits are plotted slightly closer together. Otherwise the lightcurves are as for Fig.~1}
\end{figure}

\begin{figure}
\hspace*{-12mm}\includegraphics[width=10cm]{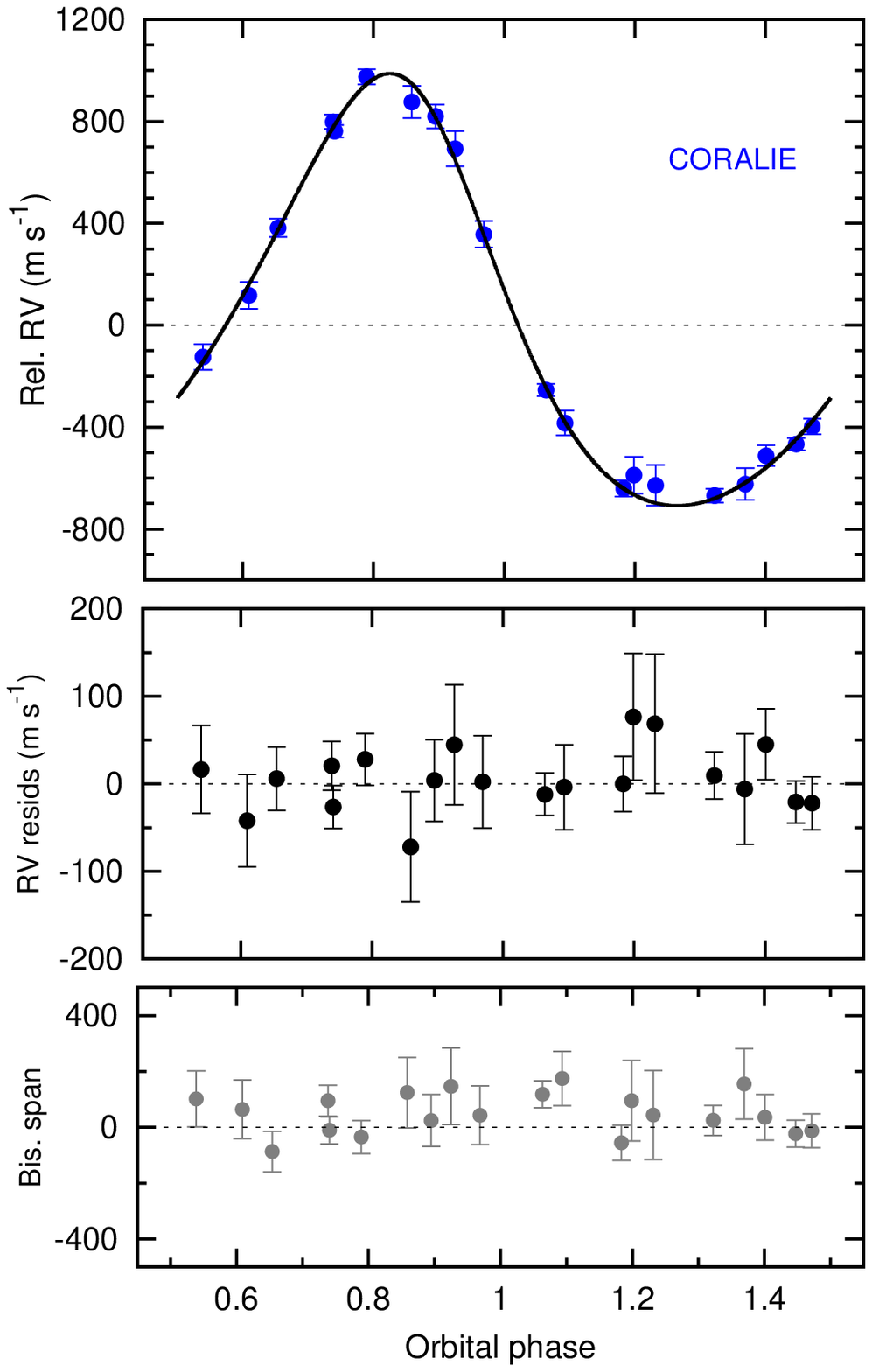}\\ [-2mm]
\caption{WASP-89b radial velocities (as for Fig.~2).}
\end{figure}

The fitted parameters were $T_{\rm c}$, $P$, $\Delta F$, $T_{14}$,
$b$, $K_{\rm 1}$, where $T_{\rm c}$ is the epoch of mid-transit, $P$
is the orbital period, $\Delta F$ is the fractional flux-deficit that
would be observed during transit in the absence of limb-darkening,
$T_{14}$ is the total transit duration (from first to fourth contact),
$b$ is the impact parameter of the planet's path across the stellar
disc, and $K_{\rm 1}$ is the stellar reflex velocity
semi-amplitude. The transit lightcurves lead directly to stellar
density but one additional constraint is required to obtain stellar
masses and radii, and hence full parametrisation of the system.  Here
we use the calibrations presented by Southworth (2011), based on
masses, radii and effective temperatures of eclipsing binaries.

For each system we list the resulting parameters in Tables~2 to 4, and
plot the resulting data and models in Figures~1 to 4.  We also refer
the reader to Smith \etal\ (2012) who present an extensive analysis of
the effect of red noise in the transit lightcurves on the resulting
system parameters.

\section{Evolutionary status}
\begin{figure}
\mbox{\includegraphics[width=0.49\textwidth]{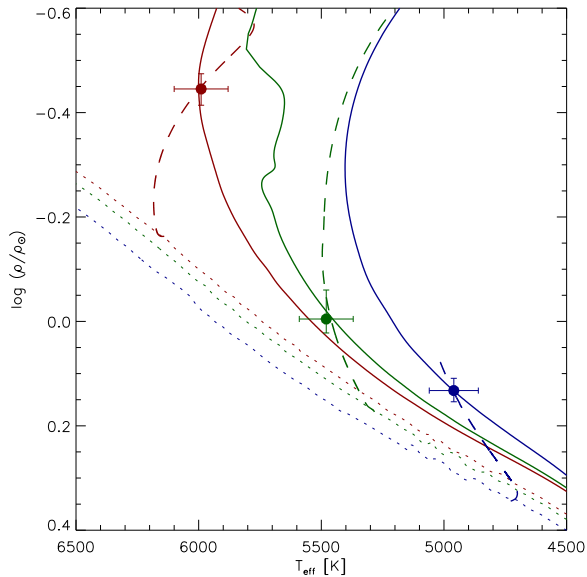}}
\caption{Mean stellar density versus effective temperature for WASP-74 (red),
WASP-83 (green) and WASP-89 (blue). The best-fit isochrones (color-coded solid lines) are at 3.7 Gyr (WASP-74), 7.1 Gyr (WASP-83) and 5.1 Gyr (WASP-89).  The  stellar evolution tracks (color-coded dashed lines) are for masses 1.31 \Msolar\ (WASP-74), 1.00  \Msolar\ (WASP-83) and 0.87  \Msolar\ (WASP-89).  Lines for each star were interpolated from our grid of
{\sc garstec} models with $\alpha_{\rm MLT}=1.78$  using the parameters given
in Table~\ref{AgeMassTable} and the appropriate value of [Fe/H] for each star.
The dotted line shows an isochrone at an age of 0.1\,Gyr at the same values of
[Fe/H].
\label{AgeMassFig}}
\end{figure}

One area where the methods of this paper differ from those of previous
WASP-South discoveries is in the comparison of the stellar parameters
to evolutionary models. Here we use the method described in detail in
Maxted, Serenelli \&\ Southworth (2014).  This uses an MCMC method to
calculate the posterior distribution for the mass and age estimates of
the star, by comparing the observed values of \rhostar, \Teff\ and
[Fe/H] to a grid of stellar models.  The stellar models were
calculated using the {\sc garstec} stellar evolution code (Weiss \& Schlattl (2008) and the methods used to calculate the
stellar model grid are described in Serenelli \etal\ (2013) The
results of this Bayesian analysis are given in
Table~\ref{AgeMassTable} and are shown in Fig.~\ref{AgeMassFig}.

\begin{table}
 \caption{Bayesian mass and age estimates for the host stars  Columns 2 and 3 give the maximum-likelihood estimates of the age and
mass, respectively.  Columns 4 and 5 give the mean and standard deviation of the 
posterior age and mass distribution, respectively.
\label{AgeMassTable}}
 \begin{tabular}{@{}lrrrr}
\hline
  \multicolumn{1}{@{}l}{Star} &
  \multicolumn{1}{l}{$\tau_{\rm b}$} &
  \multicolumn{1}{l}{$M_{\rm b}$} &
  \multicolumn{1}{l}{$\langle \taustar \rangle$ }  &
  \multicolumn{1}{l}{$\langle \Mstar \rangle$ }\\
  \multicolumn{1}{l}{} &
  \multicolumn{1}{l}{[Gyr]} &
  \multicolumn{1}{l}{[\Msolar]} &
  \multicolumn{1}{l}{[Gyr]}  &
  \multicolumn{1}{l}{[\Msolar]}\\
\hline
 \noalign{\smallskip}
WASP-74     & 3.5 & 1.32 & $ 3.7 \pm 0.9$&$ 1.31 \pm 0.06 $\\
WASP-83     & 6.3 & 1.01 & $ 7.1 \pm 2.9$&$ 1.00 \pm 0.05 $\\
WASP-89     &14.9 & 0.81 & $12.5 \pm 3.1$&$ 0.84 \pm 0.04 $\\
WASP-89$^a$ & 1.8 & 0.91 & $ 5.1 \pm 3.3$&$ 0.87 \pm 0.04 $\\
 \noalign{\smallskip}
\hline
\multicolumn{5}{@{}l}{$^a$ Assuming $\alpha_{\rm MLT}=1.22$}
 \end{tabular}   
 \end{table}

\section{WASP-74}
WASP-74 is a $V$ = 9.7, F9 star with a metallicity of [Fe/H] = +0.39
$\pm$ 0.13.  The transit analysis gives a mass and radius of 1.48
$\pm$ 0.12 M$_{\rm \odot}$ and 1.64 $\pm$ 0.05 R$_{\rm \odot}$.  The
transit $\log g_{*}$ of 4.20 $\pm$ 0.02 compares to a spectroscopic
$\log g_{*}$ of 4.39 $\pm$ 0.07.  The evolutionary comparison
(Fig.~\ref{AgeMassFig}) suggests an evolved star with an age of 3.7
$\pm$ 0.9 Gyr and a lower mass of 1.31 $\pm$ 0.06 M$_{\rm \odot}$.
The gyrochronological and lithium age estimates are lower at
$2.0^{+1.6}_{-1.0}$ Gyr and \sqiggt 2 Gyr respectively.

The planet, WASP-74b, is a relatively typical hot Jupiter in a 2-d
orbit, having a mass of 0.95 $\pm$ 0.06 M$_{\rm Jup}$ and a moderately
bloated radius of 1.56 $\pm$ 0.06 R$_{\rm Jup}$.

\section{WASP-83}
WASP-83 is a fainter G8 star of $V$ = 12.9, with a metallicity of
[Fe/H] = +0.29 $\pm$ 0.12. The spectroscopic $\log g$ of 4.34 $\pm$
0.08 is compatible with the transit $\log g$ of 4.44
$^{+0.02}_{-0.04}$.  The mass of 1.11 $\pm$ 0.06 M$_{\rm \odot}$ from
the transit analysis is in line with the evolutionary estimate of 1.00
$\pm$ 0.05 M$_{\rm \odot}$.  The evolutionary age of 7.1 $\pm$ 2.9 Gyr
is in line with the lithium age of \sqiggt 5 Gyr while the \vsini\ of $<$\,0.5 \kmps\ also implies an old star. 

The planet, WASP-83b, has a mass of 0.30 $\pm$ 0.03 M$_{\rm Jup}$,
matching that of Saturn, and a moderately bloated radius of 1.04
$^{+0.08}_{-0.04}$ R$_{\rm Jup}$.  It is very similar to WASP-21b
(Bouchy \etal\ 2010), which has a similar mass (0.3 M$_{\rm Jup}$), is
also bloated (1.2 R$_{\rm Jup}$), and also has a 4-d orbit around a G
star.

\section{WASP-89}
With a magnitude of $V$ = 13.1, WASP-89 is among the faintest planet-hosts found by WASP-South, but is among the more interesting systems. The
spectroscopy (with a low S/N owing to the faintness) reports it as a
K3 star with $\log g$ = 4.31 $\pm$ 0.16 and a mass of 0.88 $\pm$ 0.08
M$_{\rm \odot}$. The transit analysis gives $\log g$ = 4.52 $\pm$
0.02, with a mass of 0.92 $\pm$ 0.08 M$_{\rm \odot}$.

The initial Bayesian evolutionary analysis of WASP-89 (Table~\ref{AgeMassTable}) gave an
age of 12.5 $\pm$ 3.1 Gyr, which would likely make it older than the
Galactic disk. This raises the
possibility that this star is affected by the ``radius anomaly''
observed in many other late-type stars, particularly those like
WASP-89 that show signs of magnetic activity (Hoxie 1973; Popper 1997; L{\'o}pez-Morales 2007; Spada \etal\ 2013). It has been proposed that this is due to the
reduction in the efficiency of energy transport by convection, a
phenomenon that can be approximated by reducing the mixing length
parameter used in the model (Feiden \& Chaboyer 2013; Chabrier, Gallardo \& Baraffe 2007). The mixing length parameter used to calculate
our model grid is $\alpha_{\rm MLT}=1.78$. With this value of
$\alpha_{\rm MLT}$ {\sc garstec} reproduces the observed properties of
the present day Sun assuming that the composition is that given by Grevesse \&\ Sauval (1998), the overall initial metallicity is Z =
0.01876, and the initial helium abundance is $Y=0.269$.  There is
currently no objective way to select the correct value of $\alpha_{\rm
  MLT}$ for a magnetically active star other than to find the range of
this parameter that gives plausible results. Accordingly, we also
calculated a Markov chain for the observed parameters of WASP-89 using
stellar models with $\alpha_{\rm MLT}=1.22$, for which value we find
$p(\taustar < 10\,{\rm Gyr}) = 0.91$. By comparing the results for
WASP-89 with the two values of $\alpha_{\rm MLT}$ we arrive at a mass
of $0.85\pm0.05$\,\Msolar\ with the age being indeterminate. This is
then compatible with the masses from the spectral analysis and the
transit analysis.

\begin{table}
 \caption{Periodogram analysis of the WASP lightcurves for WASP-89. Observing
dates are JD -- 2450\,000,  $N$pts is the number of data points, Amp is the semi-amplitude (in magnitudes) of the best-fit sine wave at the period $P$ found
in the periodogram with false-alarm probability FAP.
\label{ProtTable}}
 \begin{tabular}{@{}lrrrrr}
\hline
  \multicolumn{1}{@{}l}{Year} &
  \multicolumn{1}{l}{Dates} &
  \multicolumn{1}{l}{$N$pts} &
  \multicolumn{1}{l}{$P$ [d]} &
  \multicolumn{1}{l}{Amp} &
  \multicolumn{1}{r}{FAP}\\
 \noalign{\smallskip}
\hline
2008 & 4622--4752 &  6648   & 20.69  &  0.007  & 0.006  \\
2009 & 4970--5116 &  5589   & 10.46  &  0.006  & 0.072  \\
2011 & 5691--5858 &  4072   & 19.65  &  0.014  & $<0.001$ \\
2012 & 6053--6107 &  1430   & 19.57  &  0.010  & 0.004 \\
 \noalign{\smallskip}
\hline
 \end{tabular}   
 \end{table}    

\begin{figure}[t]
\mbox{\includegraphics[width=0.49\textwidth]{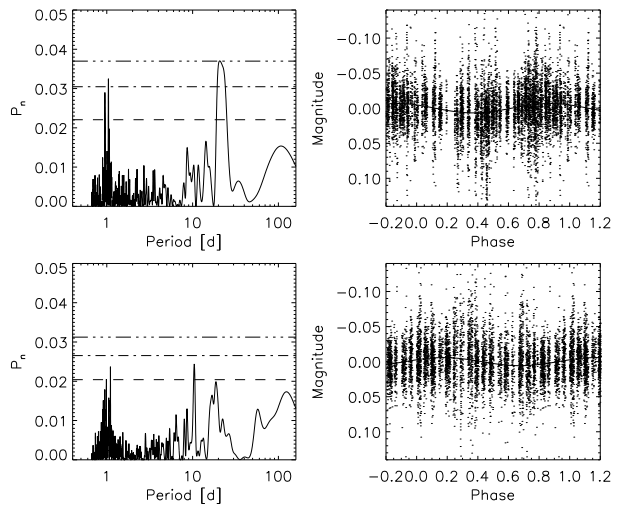}}
\caption{Periodograms of the WASP lightcurves for WASP-89 obtained in 2008 (top  left) and 2009 (bottom left). Horizontal lines indicate false-alarm probability levels of 0.1, 0.01 and 0.001.  Top right shows the 2008 data folded on 20.69 d; bottom right the 2009 data folded in 10.46 d.\label{ProtFig}}
\end{figure}

\subsection{Magnetic activity}
WASP-89 shows clear evidence of magnetic activity in the form of a
rotational modulation and through star-spots during transit. Three
years of WASP-South data all show a $\sim$\,1 per cent modulation at a
period near 20 d, and the fourth shows a modulation at half that (10
d), presumably the first harmonic of the rotational period caused by a
more complex spot pattern (Table~6, Fig.~7). The average from four
different years of WASP-South data is a rotational period of $P_{\rm
  rot} = 20.2 \pm 0.4$\,d.  This, together with our value for the
stellar radius, implies a value of $V_{\rm rot} = 2.2 \pm 0.1$\,\kms,
which compares to the spectroscopic $V_{\rm rot}\sin I$ estimate of
$2.5 \pm 0.9$ \kms.  This is consistent with the WASP-89's spin axis
being at 90$^{\circ}$ to us.

The transit lightcurves from TRAPPIST and EulerCAM show clear star
spots, visible as bumps in the transit profile, most clearly at phase
0.997 in the EulerCAM lightcurve from the 02 Oct 2012 (Fig.~4).  This raises
the issue of whether we are treating the lightcurves correctly in our
MCMC analysis (See the discussion in, e.g., Oshagh \etal\ 2013).
  When a planet transits a spot we see a slight
brightening, and including such data will cause the fitted transit to
be shallower.  However, any spots that are not transited but still
present will do the opposite, causing the transit to be deeper.  Thus
excluding bumps caused by transited spots would introduce a
bias. Without more information on the extent of spottedness there is
no secure way of dealing with this.  The rotation modulation suggests
that the difference between different faces of the star is of order 1
per cent of the brightness, which is comparable to other
uncertainties.  We have thus chosen to simply combine all the
lightcurves in the analysis, effectively averaging over any spots
present.

In principle one can use transits of star spots to deduce the orbital
alignment (e.g.~Tregloan-Reed, Southworth \&\ Tappert 2013). The
TRAPPIST and EulerCAM lightcurves from 12 Sep 2012 are of the same
transit, as are those from 02 Oct 2012, the latter being 6 orbital
cycles (20.1 d) later.  There appears to be a spot at phase 0.992 in
the 12 Sep lightcurve and a spot at 0.997 in the 02 Oct lightcurve.
This could be the same spot being transited one stellar rotation
later, which is unlikely unless the planet's orbit is aligned with the
stellar rotation (or anti-aligned).

If it is the same spot, the difference in the phase of the spot
transit implies that the star had rotated by 1.07 cycles (or 0.93
cycles), which translates to a rotation period of 18.8 $\pm$ 0.3 d (or
21.7 $\pm$ 0.3 d). This is slightly different from the value of 20.2 $\pm$
0.4 d from the WASP data, but the discrepancy might be accounted for
by differential rotation. 

Thus, we conclude that WASP-89 rotates with a 20-d period and is
magnetically active, and that there are indications that the
planetary orbit is aligned or anti-aligned. However, we need more
extensive star-spot observations and observations of the
Rossiter--McLaughin effect to be sure.

\subsection{A massive planet in an eccentric orbit}
WASP-89b has a mass of 5.9 $\pm$ 0.4 M$_{\rm Jup}$ and is in a 3.356-d
orbit with an eccentricity of 0.19 $\pm$ 0.01.  It thus joins a small
number of massive planets in short-period, eccentric orbits, of which
the most similar are XO-3b (12 M$_{\rm Jup}$, 3.2 d, $e$ = 0.26;
Johns-Krull \etal\ 2008), HAT-P-2b (8.7 M$_{\rm Jup}$, 5.6 d, $e$ =
0.52; Bakos \etal\ 2007) and HAT-P-21b (4.0 M$_{\rm Jup}$, 4.1 d, $e$
= 0.23; Bakos \etal\ 2011).

It is worth noting, though, that those three planets orbit stars of
spectral type F5, F8 and G3, respectively.  WASP-89 is the first known
K star hosting a massive planet in a short-period eccentric orbit ($M
> 1$ M$_{\rm Jup}$; $P < 6$ d; $e > 0.1$).  The magnetic activity of
WASP-89 could be related to the hosting of a massive, short-period
planet, since magnetic activity might be enhanced in hot-Jupiter hosts
(e.g.~Poppenhaeger \&\ Wolk 2014).    

Planets in eccentric, short-period orbits are of particular
interest in that their rotation cannot be fully phase-locked to their orbit,
and so they must experience large differences in radiative forcing around
the orbit. Thus they can tell us about the dynamics of giant-planet
atmospheres (e.g.~Wong \etal\ 2014 and references therein).

The usual explanation for the occurrence of such eccentric orbits in
short-period hot Jupiters is that they are moved in inwards by a
process of ``high eccentricity migration'', followed by
circularisation (e.g. Rasio \&\ Ford 1996; Fabrycky \&\ Tremaine 2007;
Naoz \etal\ 2011; Socrates \etal\ 2012a).

The circularisation timescale can be estimated from (Adams \& Laughlin
2006, eqn 3):

\begin{eqnarray*}
\tau_{\rm cir}\ \approx &\ 1.6~{\rm Gyr} \times \left(\frac{Q_{\rm P}}{10^6}\right) \times \left(\frac{M_{\rm P}}{M_{\rm Jup}}\right) \times \left(\frac{M_*}{\Msolar}\right)^{-3/2} \nonumber \\  & \times \left(\frac{R_{\rm P}}{R_{\rm Jup}}\right)^{-5} \times \left(\frac{a}{0.05~{\rm AU}}\right)^{13/2}
\end{eqnarray*}

The value of the quality factor, ${Q_{\rm P}}$, is unclear, but if we
take it as 10$^{5}$ (e.g.\ Socrates, Katz \&\ Dong 2012b), then we
obtain for WASP-89b a circularisation timescale of $\sim$\,2 Gyr.
Here, the large mass of the planet prevents circularisation in less than a
Gyr despite the short orbit.  This timescale is in line with the
gyrochronological age of the host star, and thus the fact that the
planet has not circularised is consistent.  Tidal damping of
eccentricity is expected to occur faster than damping of obliquity or
inwards orbital decay (Matsumura, Peale \&\ Rasio 2010), and thus we
would expect the current values of these properties to be direct products of the
high-eccentricity migration.

\section*{Acknowledgements}
WASP-South is hosted by the South African Astronomical Observatory and
we are grateful for their ongoing support and assistance. Funding for
WASP comes from consortium universities and from the UK's Science and
Technology Facilities Council. The Euler Swiss telescope is supported
by the Swiss National Science Foundation. TRAPPIST is funded by the
Belgian Fund for Scientific Research (Fond National de la Recherche
Scientifique, FNRS) under the grant FRFC 2.5.594.09.F, with the
participation of the Swiss National Science Fundation (SNF). This
paper includes observations made with the RISE photometer on the 2.0-m
Liverpool Telescope under PATT program PL14A10. The Liverpool
Telescope is operated on the island of La Palma by Liverpool John
Moores University in the Spanish Observatorio del Roque de los
Muchachos of the Instituto de Astrofisica de Canarias with financial
support from the UK Science and Technology Facilities Council.
M. Gillon and E. Jehin are FNRS Research Associates.  A.H.M.J. Triaud
is a Swiss National Science Foundation Fellow under grant
P300P2-147773. L. Delrez acknowledges the support of the F.R.I.A. fund
of the FNRS.

\clearpage



\begin{table}
\caption{Radial velocities.\protect\rule[-1.5mm]{0mm}{2mm}} 
\begin{tabular}{cccr} 
\hline 
BJD\,--\,2400\,000 & RV & $\sigma$$_{\rm RV}$ & Bisector \\
(UTC)  & (km s$^{-1}$) & (km s$^{-1}$) & (km s$^{-1}$)\\ [0.5mm] \hline
\multicolumn{4}{l}{{\bf WASP-74:}}\\  
55795.60542 &	$-$15.8781 &	0.0045 &	$-$0.0067  \\
55796.64971 &	$-$15.6394 &	0.0047 &	$-$0.0094  \\
55802.72044 &	$-$15.6772 &	0.0056 &	0.0026	  \\
55820.55636 &	$-$15.7453 &	0.0042 &	0.0028	  \\
55823.52316 &	$-$15.8463 &	0.0060 &	0.0014	  \\
55824.54772 &	$-$15.6601 &	0.0045 &	$-$0.0213  \\
55826.57775 &	$-$15.6553 &	0.0049 &	$-$0.0019  \\
55827.50202 &	$-$15.8600 &	0.0054 &	$-$0.0132  \\
55828.57525 &	$-$15.6596 &	0.0051 &	$-$0.0147  \\
55829.63956 &	$-$15.8690 &	0.0044 &	0.0028  \\
55830.50481 &	$-$15.6975 &	0.0050 &	$-$0.0227  \\
55832.53917 &	$-$15.7364 &	0.0042 &	0.0048  \\
55834.53461 &	$-$15.7850 &	0.0042 &	$-$0.0112  \\
55835.53651 &	$-$15.7357 &	0.0050 &	$-$0.0019  \\
55851.58947 &	$-$15.7856 &	0.0048 &	0.0015	  \\
56049.91441 &	$-$15.9043 &	0.0041 &	$-$0.0147  \\
56103.75557 &	$-$15.8386 &	0.0039 &	$-$0.0158  \\
56150.65595 &	$-$15.8523 &	0.0051 &	$-$0.0261  \\
56152.55759 &	$-$15.8601 &	0.0052 &	$-$0.0277  \\
56212.54690 &	$-$15.8647 &	0.0042 &	0.0007	  \\ [0.5mm] 
\multicolumn{4}{l}{{\bf WASP-83:}}\\  
55626.85323 &	31.5152 & 	0.0160 &	$-$0.0290 \\
55648.77839 &	31.5828 & 	0.0120 &	0.0340	 \\
55651.73866 &	31.4915 & 	0.0162 &	$-$0.0095 \\
55675.76117 &	31.5583 & 	0.0176 &	0.0029	 \\
55676.66336 &	31.5096 & 	0.0132 &	0.0412	 \\
55677.75049 &	31.5340 & 	0.0123 &	0.0106	 \\
55679.65881 &	31.5543 & 	0.0109 &	0.0237	 \\
55680.67015 &	31.5312 & 	0.0148 &	0.0003	 \\
55682.61842 &	31.5544 & 	0.0081 &	$-$0.0127 \\
55683.69704 &	31.5697 & 	0.0093 &	0.0069	 \\
55684.70318 &	31.5667 & 	0.0100 &	0.0015	 \\
55707.63183 &	31.5492 & 	0.0114 &	0.0389	 \\
55711.55909 &	31.4847 & 	0.0194 &	0.0092	 \\
55721.53003 &	31.5056 & 	0.0117 &	$-$0.0133 \\
55767.53761 &	31.5751 & 	0.0141 &	0.0194	 \\
55768.48367 &	31.5820 & 	0.0133 &	0.0048	 \\
55769.49626 &	31.5754 & 	0.0092 &	0.0221	 \\
55952.79597 &	31.5772 & 	0.0083 &	0.0112	 \\
55958.80682 &	31.5390 & 	0.0089 &	$-$0.0186 \\
55978.72739 &	31.5360 & 	0.0131 &	0.0089	 \\
55981.69825 &	31.5742 & 	0.0148 &	$-$0.0285 \\
55982.78430 &	31.5790 & 	0.0094 &	$-$0.0113 \\
55983.85735 &	31.5348 & 	0.0091 &	$-$0.0091 \\
55984.67552 &	31.5218 & 	0.0097 &	0.0191	 \\
55985.69305 &	31.5433 & 	0.0097 &	0.0102	 \\
56030.67402 &	31.5659 & 	0.0136 &	0.0052	 \\
56038.67878 &	31.5257 & 	0.0149 &	$-$0.0179 \\
56336.84838 &	31.5461 & 	0.0115 &	0.0185	 \\ [0.5mm]
\hline
\multicolumn{4}{l}{Bisector errors are twice RV errors} 
\end{tabular} 
\end{table} 

\begin{table}
\begin{tabular}{cccr} 
\hline 
BJD\,--\,2400\,000 & RV & $\sigma$$_{\rm RV}$ & Bisector \\
(UTC)  & (km s$^{-1}$) & (km s$^{-1}$) & (km s$^{-1}$)\\ [0.5mm] \hline
\multicolumn{4}{l}{{\bf WASP-89:}}\\  
55685.90591 &	21.8497 &	0.0245 &	$-$0.0109 \\
56123.72535 &	20.4472 &	0.0315 &	$-$0.0556 \\
56124.69378 &	20.6905 &	0.0303 &	$-$0.0124 \\
56125.75998 &	22.0629 &	0.0296 &	$-$0.0349 \\
56126.68173 &	20.8330 &	0.0242 &	0.1185	 \\
56136.84966 &	20.7044 &	0.0486 &	0.1745	 \\
56150.73985 &	20.4600 &	0.0794 &	0.0443	 \\
56151.77120 &	20.9629 &	0.0503 &	0.1013	 \\
56154.55728 &	20.4647 &	0.0630 &	0.1552	 \\
56158.71898 &	21.2050 &	0.0527 &	0.0643	 \\
56159.55775 &	21.9642 &	0.0629 &	0.1241	 \\
56159.78044 &	21.7811 &	0.0685 &	0.1462	 \\
56181.67023 &	20.6210 &	0.0240 &	$-$0.0228 \\
56182.64854 &	21.8869 &	0.0278 &	0.0953	 \\
56184.60865 &	20.4188 &	0.0268 &	0.0246	 \\
56186.53071 &	21.9080 &	0.0464 &	0.0243	 \\
56187.54974 &	20.5000 &	0.0723 &	0.0951	 \\
56203.56053 &	21.4453 &	0.0526 &	0.0432	 \\
56212.57303 &	21.4703 &	0.0364 &	$-$0.0872 \\
56419.82154 &	20.5757 &	0.0407 &	0.0355	 \\ [0.5mm] 
\hline
\multicolumn{4}{l}{Bisector errors are twice RV errors} 
\end{tabular} 
\end{table} 

\end{document}